\documentclass[preprint,preprintnumbers,amsmath,amssymb]{revtex4}

\usepackage{color}
\usepackage{amsmath,graphicx,amsfonts,mathrsfs,amssymb}
\usepackage{amsmath,epsfig}
\usepackage{amssymb,amsfonts}
\usepackage{latexsym}
\usepackage{epsfig}
\newbox\pippobox
\def\be{\begin{equation}}
\def\ee{\end{equation}}
\def\bea{\begin{eqnarray}}
\def\eea{\end{eqnarray}}

\def\ee           {{\rm e}}

\newcommand{\beq}{\begin{equation}}
\newcommand{\eeq}{\end{equation}}
\newcommand{\beqa}{\begin{eqnarray}}
\newcommand{\eeqa}{\end{eqnarray}}
\newcommand{\beqar}{\begin{eqnarray*}}
\newcommand{\eeqar}{\end{eqnarray*}}

\renewcommand{\eqref}[1]{(\ref{#1})}

\newcommand{\ren}{R\'enyi\ }

\catcode`\@=12

\begin{document}

\title{Holographic \ren Entropy by Generalized Entropy method\\ \vspace{2cm}
\vspace{1cm}}

\author{Wu-zhong Guo}\email{wuzhong@itp.ac.cn}
\author{Miao Li$^+$}
\affiliation{Kavli Institute for Theoretical Physics, Key Laboratory
of Frontiers in Theoretical Physics, Institute of Theoretical
Physics, Chinese Academy of Sciences, Beijing 100190\\
$^+$Institute of Astronomy and Space Science, Sun Yat-Sen University, Guangzhou 510275,
People's
Republic of China \vspace{2cm}}

\begin{abstract}
In this paper we use the method of generalized gravitational entropy in \cite{Lewkowycz:2013nqa} to construct the
dual bulk geometry for a spherical entangling surface, and calculate the \ren entropy with the dual bulk gravity
theory being either Einstein gravity or  Lovelock gravity, this approach is closely related to that in \cite{Casini:2011kv}.
For a general entangling surface we derive the area law of entanglement entropy. The area law is closely related with
the local property of the entangling surface.

\end{abstract}

\maketitle

\section{Introduction}
The entanglement entropy is an useful quantity to characterize the non-local
property of a quantum field theory. Given a subsystem A, the degree of
freedom (DOF) that is only accessible to an observer in A can be described
by the density matrix $\rho_A$, which is defined by taking trace over the DOFs outside
the region $A$, i.e., $\rho_A=tr_{\bar A}\rho$, where $\bar A$ is the complementary
part of the subsystem A, $\rho$ is the density matrix for the whole system.
The entanglement entropy is defined as the von Neumann entropy of the reduced density
$\rho_A$: $S_A=-tr(\rho_A\log \rho_A)$, its holographic calculation for a CFT with a gravity dual was proposed in \cite{Ryu:2006bv}. The
entanglement entropy in d-dimensional boundary theory for a spatial region A on a constant time slice is
calculated by
\begin{eqnarray}\label{HolographicFormula}
S_A=min[\frac{s(A)}{4G}],
\end{eqnarray}
where $s(A)$ is the area of the codimension-2 surface in the bulk spacetime which is homologous to
the boundary of spatial region $A$.\\
The \ren entropy can also be used as a quantity to describe the entanglement property. Given the reduced
normalized density matrix $\rho_A$, the entanglement \ren entropy is a series of quantities defined as
\begin{eqnarray}\label{RenyiFormula}
S_n=\frac{\log tr[\rho_A^n]}{1-n},
\end{eqnarray}
 where $0<n<\infty$. The entanglement \ren entropy contains much more information about entanglement.
 In fact, knowing the \ren entropy $S_n$ only for all integers $n$ we can calculate the full eigenvalues of
 the reduced density matrix $\rho_A$. The standard way to calculate the entanglement \ren entropy is using
 the ``replica trick'', requiring evaluating the partition function on a singular manifold. The exact
 results are only known for some simply cases. The entanglement entropy is just the entanglement \ren entropy
 by taking the limit $n\to 1$ in (\ref{RenyiFormula}). When considering the holographic approach for the field
 theory living on the boundary with a gravity dual, it seems that we need to construct the dual bulk geometry for the
 singular manifold after performing the replica trick by using the dual relation to calculate the partition function. In \cite{Fursaev:2006ih} the
 author attempts to use the singular bulk space which has
 the required boundary singular manifold structure to prove the holographic entanglement entropy proposal
 (\ref{HolographicFormula}). However, it is pointed out in \cite{Headrick:2010zt} that the approach in \cite{Fursaev:2006ih} would produce
 an incorrect result of the bulk solution, and the entanglement \ren entropy is also independent of $n$. \\
 On the other hand, in \cite{Casini:2011kv} the authors find
 the equivalence between the reduced density matrix $\rho_A$ of a ball in d-dimensional flat spacetime
 and the thermal density matrix in background $R\times H^{d-1}$ for CFT living on the boundary. This approach eliminates a singular boundary. One can calculate the thermal entropy in the bulk side, which is a topological black hole with hyperbolic horizon. The entanglement \ren entropy is also calculable following the same
 step, which is consistent with the known result in 2-dimensions \cite{Hung:2011nu}. But it is not easy to generalize this method to arbitrary entangling surface. \\
A recent paper \cite{Lewkowycz:2013nqa} establishes a method, named generalized gravitational entropy, to calculate entropy of an Euclidean black hole solution
 without a Killing vector. The entanglement entropy of  an arbitrary spatial subregion for the CFT with a gravity dual is an example
 of this general case by the gauge/gravity duality. This argument would lead to the proposal (\ref{HolographicFormula}) beyond the spherical
shape \cite{Casini:2011kv}\footnote{The subsequent papers \cite{Chen:2013qma}\cite{Bhattacharyya:2013jma} point out there are some mismatches for higher derivative gravity theory, but the mismatch is removed by the paper \cite{Dong:2013qoa}}.  In this paper we first discuss the proof for the spherical entangling surface case, and use the result to
calculate the entanglement \ren entropy, the result is consistent with that in \cite{Hung:2011nu}. We also notice the analogy
between these two approaches. The generalized gravitational entropy \cite{Lewkowycz:2013nqa} provides us a method to study the entanglement entropy for a general entangling surface. We find that the area law of the entanglement entropy is closely related with the local property of the entangling surface. \\

This paper is organized as follows. In section \uppercase\expandafter{\romannumeral2} we briefly review the generalized gravitational entropy method and construct the dual
bulk geometry for the spherical entangling surface. By using this result we study the entanglement \ren entropy with a spherical entangling surface in section \uppercase\expandafter{\romannumeral3}. We also notice the relation between the generalized gravitational entropy method and the previous approach \cite{Casini:2011kv}. In section \uppercase\expandafter{\romannumeral4} the area law of the entanglement entropy is derived for a general entangling surface. Section \uppercase\expandafter{\romannumeral5} contains some discussion about the gravity theory without a dual field theory on the boundary and the conclusion.

\section{The Generalized Gravitational Entropy Method}
The argument of the generalized gravitational entropy begins with the assumed underlying density matrix $\rho$ in the full quantum gravity
theory. In the classical approximation, the trace of $\rho$ can be explained as Euclidean evolution of the gravity system on a ``time'' $\tau$
circle with period $2\pi$. The trace of $\rho^n$ is considered as the evolution on a circle with period $2n\pi$. Then the entropy of
the system can be calculated by the ``replica trick''. The boundary condition of the system does not necessarily have a $U(1)$ symmetry
along the time circle. We get $\log tr\rho$ by the saddle point approximation, which is the on-shell Euclidean action
with the period of the time circle being $2\pi$. $\log tr\rho^n$ is also the on-shell Euclidean action, with a new Einstein solution.
The new solution is for the spacetime with the time circle $\tau\sim \tau+2n\pi$, while the boundary condition remains the same. The entropy
would be one quarter of the area of a special codimension-2 surface where the time circle shrink to zero. It is shown in \cite{Lewkowycz:2013nqa} that the special surface satisfies the minimal area condition, see \cite{Lewkowycz:2013nqa} for more details.\\

The above result is  related to the holographic entanglement entropy of subregion $A$ in field theory using the gauge/gravity correspondence.
The reduced density matrix $\rho_A$ can be mapped to the corresponding density matrix $\rho$ for the quantum theory in the bulk. The ``replica trick''
is actually the same as we stated above. The entanglement entropy is equal to the gravitational entropy in the bulk, which is related to the special
codimension-2 surface. The remaining work is to prove that this codimension-2 surface is homologous to the boundary of subregion $A$. The holographic entanglement entropy proposal (\ref{HolographicFormula}) can be derived then. In the next two subsections we will derive the analytical results for the infinite plane and the sphere.
\subsection{An infinite plane}
The standard representation of the density matrix $\rho_A$ of a spatial subregion $A$ in a field theory is the path integral over
the field configurations on spacetime with a cut on the entangling surface $\partial A$. The entangling surface is defined as the boundary of the
subregion $A$ on a constant slice of time. $tr \rho^n$ is the path integral on the spacetime which has a conical defect with a $2n\pi$
at $\partial A$. The key point of the argument in \cite{Lewkowycz:2013nqa} is that one can use a suitable
conformal transformation to ensure that the ``time'' circle around $\partial A$ does not shrink to zero size. Here  ``time'' is usually not the original time coordinate in the field theory. Now it is safe to use the ``replica trick'' without worrying about the appearance of singularity. \\

 Assuming we have a conformal field theory living on the d-dimensional Euclidean space $R^d$. The subregion $A$ is defined by $x>0$. The entangling
 surface is $t=0, x=0$. We denote the reduced density matrix by $\rho_A$. Now, as is shown in \cite{Lewkowycz:2013nqa}, one can redefine
 coordinates $r,\tau$, with $t=r\sin\tau$ and $x=r\cos\tau$. The metric of
 our spacetime is
\begin{eqnarray}\label{TransformationPlane}
 ds^2=\Omega^2 \Big(d\tau^2+\frac{dr^2+\sum_{i=1}^{d-2} dx_i^2}{r^2}\Big),
\end{eqnarray}
where $\Omega=r$. The original spacetime is mapped to the manifold $R\times H^{d-1}$ by a conformal transformation eliminating the
factor $\Omega^2$.\\
The subregion $A$ corresponds to region $\tau=0$,$r>0$, and its complementary part $\bar A$ corresponds
to $\tau=\pi$,$r>0$  on the new manifold. $r=0$ defines the entangling surface. The reduced density matrix $\rho_A$ is mapped to an operator
$\tilde\rho_A$ on the new manifold by a unitary transformation, $\tilde\rho_A$=$U$ $\rho_A$ $U^{-1}$. The new density
matrix $\tilde \rho_A$ is much easier to get if the conformal field theory has a dual gravity theory in the bulk. The manifold is the boundary
of $AdS_{d+1}$ with the radius $L=1$. According to AdS/CFT corresponding $\tilde \rho_A$ can be mapped to the density matrix of the bulk, which is the topological black hole with a hyperbolic horizon. The entanglement entropy is equal to $\frac{A}{4G}$, where $A$ is the area of the horizon. The remaining work is to show that the horizon is homologous to the entangling surface $r=0$, the details are presented in appendix $A$.

\subsection{A sphere}
The entanglement entropy of a sphere is another simple case in which we can get the analytic
result by the generalized entropy method. In this section we give the construction
of the proof, and the result will be used in section \uppercase\expandafter{\romannumeral3} to study the entanglement \ren
entropy. The general idea is same as the infinite plane case.\\

We assume that the CFT lives on the flat d-dimensional Euclidean space $R^d$, with metric
\begin{eqnarray}\label{Sphere}
ds^2=dt^2+dr^2+r^2d\Omega_{d-2}^2.
\end{eqnarray}
 Our spatial subregion $A$ is a ball with the entangling
surface being $S^{d-2}$ with radius $R=1$ on a constant Euclidean time slice, i.e., $t=0$ and $r=R$.
The main work is to find new manifold which has the property that the ``time'' coordinate do not
shrink to zero at the entangling surface. And the new manifold should be conformal to $R^d$. We
find the following coordinate transformation:
\begin{eqnarray}\label{Transformation}
t=\frac{\sin\tau}{\cosh u+\cos\tau}\quad \quad and \quad \quad r=\frac{\sinh u}{\cosh u +\cos\tau},
\end{eqnarray}
where the new coordinate $0<u<\infty$ and $\tau$ has the period $\tau\sim \tau+2\pi$. The subregion $A$, $t=0$ and $r<1$,
is mapped to $\tau=0$ and $0<u<\infty$, its complementary part is mapped to $\tau=\pi$ and $0<u<\infty$.
Metric (\ref{Sphere}) is transformed to
\begin{eqnarray}\label{BoundarySphere}
ds^2=\Omega^2\Big(d\tau^2+du^2+\sinh^2ud\Omega_{d-2}^2\Big),
\end{eqnarray}
where $\Omega=(\cosh u+\cos \tau)^{-1}$,  divergent on the entangling surface. Factor $\Omega^2$ can be eliminated by conformal transformation,
 the new manifold is also $S^1\times H^{d-1}$. The n-th power of the density matrix on $S^1\times H^{d-1}$ is
 equal to the path integral of fields on  manifold $S^n\times H^{d-1}$. The boundary has a well-defined dual bulk geometry. By using the argument in the introduction,
 the entanglement entropy is obtained, given by $\frac{A}{4G}$, where $A$ is the area of the horizon of the topological black
 hole. In appendix $A$ we also show that the horizon is homologous to the entangling surface $S^{d-2}$.

\section{Holographic \ren Entropy}
Actually the method to calculate the holographic entanglement entropy in section \uppercase\expandafter{\romannumeral2}
automatically provides us a way to calculate the holographic entanglement \ren entropy for CFT theories with
gravity dual \footnote{Although the entanglement entropy is the limit
$n\to 1$ of \ren entropy ,it is not necessary to calculate the \ren entropy
in order to get the entanglement entropy. The off shell and apparent conical
singularities methods are used in \cite{Lewkowycz:2013nqa} to get the result.}.
In this section we will choose the entangling surface
to be a sphere with a radius $R=1$. As we have noted in section \uppercase\expandafter{\romannumeral2}, the reduced
density matrix $\rho _A=U^{-1}\frac{\tilde\rho_A}{tr \tilde\rho_A}U$. The trace
of $n-th$ power of $\rho _A$
\begin{eqnarray}
tr(\rho_A^n)=\frac{tr\tilde\rho_A^n}{(tr\tilde\rho_A)^n}.
\end{eqnarray}
According to the definition of \ren entropy (\ref{RenyiFormula}),
\begin{eqnarray}\label{RenyiCalculation}
S_n=\frac{1}{1-n}\Big(I(n)-n I(1)\Big),
\end{eqnarray}
where $I(n)$ is the on-shell Euclidean action with the period of Euclidean time $\tau\sim \tau +2\pi n$.
Now the problem is how to construct the bulk geometry with the period of the Euclidean time
being $2n\pi $. The boundary of the bulk geometry should be $S^n\times H^{d-1}$
as discussed in section \uppercase\expandafter{\romannumeral2}. The theory would admit
the following form of metric with the asymptotical boundary $S^n\times H^{d-1}$
\begin{eqnarray}\label{BulkGeometry}
ds^2=f(\rho)N^2d\tau^2+\frac{d\rho^2}{f(\rho)}+\rho^2(du^2+\sinh^2ud\Omega^2_{d-2}),
\end{eqnarray}
$f(\rho)$ is determined by the field equation of the bulk theory we are considering, $N$ is a constant to keep the curvature scale of hyperbolic spatial slices to be $R=1$.
$f(\rho)N^2\to \rho^2$ in the limit $\rho\to \infty$, which ensures the boundary is conformal to (\ref{BoundarySphere}).\\
The position of the event horizon, where the time circle $\tau$ shrinks to zero, of the bulk geometry (\ref{BulkGeometry}) is determined by
$f(\rho_H)=0$. Now let's define a new coordinate $x^2=f(\rho)N^2$. The metric (\ref{BulkGeometry})
becomes
\begin{eqnarray}
ds^2=x^2d\tau^2+\frac{4dx^2}{f'(\rho)^2N^2}+\rho^2(du^2+\sinh^2ud\Omega^2_{d-2}).
\end{eqnarray}
Near the horizon $\rho\to\rho_H$, $x\to 0$. The condition that the period of $\tau$ is $2n\pi$ requires
\begin{eqnarray}\label{KeyFormula}
\frac{f'(\rho_H)^2N^2}{4}=n^{-2}.
\end{eqnarray}
Equation (\ref{KeyFormula}) fixes the position of horizon $\rho_H$ and the solution of the bulk geometry.\\
As an exercise to show how this works, let us consider the case when the bulk theory is Einstein gravity firstly.

\subsection{Einstein gravity}
The $d+1$ dimensional Euclidean action of Einstein gravity is
\begin{eqnarray}\label{EinsteinAction}
I=-\frac{1}{16G}\int_Md^{d+1}x\sqrt{g}\Big(R+\frac{d(d-1)}{L^2}\Big)-\frac{1}{8G}\int_{\partial M}d^dx\sqrt{h} K,
\end{eqnarray}
where $M$ denotes the bulk, $\partial M$ is regularized at the infinity boundary $\rho=\rho_{\infty}$, the
second integral is the standard Gibbons-Hawking term. To keep with previous notation we set the curvature
scale of $AdS$ $L=1$. Using the equation of motion we get
\begin{eqnarray}\label{EinsteinSoltion}
f(\rho)=\rho^2-1-\frac{C}{\rho^{d-2}},
\end{eqnarray}
where $C$ is an integration constant. We notice that the asymptotical condition, $\lim_{\rho\to \infty}f(\rho)=\rho^2$
is satisfied for the above solution (\ref{EinsteinAction}), so $N=1$. By using $f(\rho_H)=0$
we get the relation between $C$ and position of
the horizon $\rho_H$, i.e.,  $C=\rho_H^d-\rho_H^{d-2}$. Formula (\ref{KeyFormula})  determines
the position of horizon, we get
\begin{eqnarray}\label{QuadraticEquation}
d \rho_H^2-(d-2)=2n^{-1}\rho_H,
\end{eqnarray}
when $n=1$ we have $C=0$, the solution is the hyperbolic version of $AdS$. \\

Next let us turn to (\ref{RenyiCalculation}). We must know how to deal with divergence of the Euclidean action.
For $AdS$ spacetime there is a natural way to solve the problem. The Euclidean action should contain
the following terms.
\begin{eqnarray}
I=I_{bulk}+I_{surface}+I_{ct},
\end{eqnarray}
where the surface term is regularized at the boundary of the bulk on a constant slice of $\rho$, the counter term
depends only on the curvature $\mathcal R$ and its derivatives on the induced boundary \cite{Balasubramanian:1999re}\cite{Emparan:1999pm}\cite{de Haro:2000xn}. The counter terms in this method to regularize the action are referred to as Dirichlet counter terms \cite{Kofinas:2007ns}\cite{Olea:2006vd}\cite{Kofinas:2006hr}.  The form of
counter terms are related with the dimension of bulk spacetime. It is impossible to write down the counter term
explicitly for general dimensions. Action (\ref{EinsteinAction}) is
already regularized at $\rho_{\infty}$, we need to add counter terms to cancel  divergence. For example the standard
counter term for $AdS_3$ is
\begin{eqnarray}
I_{ct}^{d=2}=\frac{1}{8\pi G}\int_{\partial M}\sqrt{h},
\end{eqnarray}
where $h$ is the determinant of the boundary $\rho=\rho_\infty$. The action is
\begin{eqnarray}
I(n;d=2)=-\frac{n V_{\Sigma}}{8G}(\rho_H^2+2),
\end{eqnarray}
The entanglement \ren entropy is
\begin{eqnarray}
S_n^{(d=2)}=\frac{V_\Sigma}{8G}(1+\frac{1}{n}),
\end{eqnarray}
the result is the same as in the field theory approach \cite{Holzhey:1994we}, as well as by the holographic method \cite{Hung:2011nu}. Another example is
$AdS_5$, the action is
\begin{eqnarray}\label{Counterterm4}
I(n;d=4)=\frac{nV_{\Sigma}}{8G}\Big(-\frac{3}{4}+\rho_H^2+\rho_H^4\Big).
\end{eqnarray}
 Then we get the entanglement \ren entropy
\begin{eqnarray}
S_n^{(d=4)}=\frac{n\pi V_{\Sigma}}{8G(n-1)}\Big(2-\rho_H^2-\rho_H^4\Big),
\end{eqnarray}
 In appendix B we also calculate the cases $d=3$ and $d=5$, the general form for arbitrary dimension is
\begin{eqnarray}
S_n= \frac{n V_\Sigma L^{d-1}}{8(n-1)G}(2-\rho_H^d-\rho_H^{d-2}).
\end{eqnarray}
We will prove this in next subsection. It is interesting to take the limit $n\to 1$, we
recover the entanglement entropy
\begin{eqnarray}
S_A=\lim_{n\to 1}S_n=\frac{V_{\Sigma}}{4G},
\end{eqnarray}
where we have used  solution  to eq.(\ref{QuadraticEquation}).

\subsection{Lovelock gravity for arbitrary dimension}

The Lovelock action in $d+1$ dimensions is \cite{Lovelock:1971yv}.
\begin{eqnarray} \label{Lovelock}
I=I_{bulk}+I_{surface}=\frac{1}{16 \text{$\pi $G}}\int _Md^{d+1}x\sqrt{-g}\sum _{p=0}^{\left[\frac{d}{2}\right]} L_{2 p} \alpha _p+c_d\int _{\partial M}d^d \text{xB}_d
\end{eqnarray}
where
\begin{eqnarray}
L_{2p}\equiv \frac{1}{2^p}\delta^{[\nu_1 \nu_2...\nu_{2p-1}\nu_{2p}]}_{[\mu_1 \mu_2...\mu_{2p-1}\mu_{2p}]}R^{\mu _1 \mu _2}{}_{\nu _1 \nu _2}\cdot \cdot \cdot R^{\mu _{2 p-1} \mu _{2 p}}{}_{\nu _{2 p-1} \nu _{2 p}},
\end{eqnarray}
$B_d$ denotes the boundary term and $c_d$ is the suitable constant, $\alpha_0=\frac{d(d+1)}{L^2}$, $\alpha_1=1$ and others are arbitrary. For later convenience we also define
\begin{eqnarray}
\lambda_p=(-1)^p\frac{(d-2)!}{(d-2p)!} \frac{\alpha_p}{L^{2p-2}},
\end{eqnarray}
with $\lambda_0=1$ and $\lambda_1=-1$. Lovelock theory admits a pure $AdS_{d+1}$ solution with the curvature scale $\tilde L=L/\sqrt{f_{\infty}}$, where $f_{\infty}$ satisfies the
following expression
\begin{eqnarray}\label{ConstraintOflambda}
\sum _{p=0}^{\left[\frac{d}{2}\right]} \lambda _p f_{\infty }^p=0,
\end{eqnarray}
we choose the root $f_\infty\to 1$ in the limit $\lambda_p\to 0$. By using the equation of motion of Lovelock gravity $f(\rho)$ should satisfy the following relation
\begin{eqnarray}\label{ConstraintLovelock}
\sum_{p=0}^{[\frac{d}{2}]}\lambda_p[1+f(\rho)]^p\rho^{d-2p}L^{2p-2}=\mu,
\end{eqnarray}
where $\mu$ is the integral constant. At the horizon $\rho=\rho_H$, $f(\rho)$ is vanishing, we get
\begin{eqnarray}\label{mu}
\mu=\sum_{p=0}^{[\frac{d}{2}]}\lambda_p\rho_{H}^{d-2p}L^{2p-2}
\end{eqnarray}
and
\begin{eqnarray}\label{Lovelockderivative}
f'(\rho_H)\sum_{p=0}^{[\frac{d}{2}]}p\lambda_p \rho_{H}^{d-2p}L^{2p-2}+\sum_{p=0}^{[\frac{d}{2}]}(d-2p)\lambda_p \rho_{H}^{d-2p-1}L^{2p-2}=0.
\end{eqnarray}

From (\ref{ConstraintLovelock}) we obtain the asymptotically behavior of $f(\rho)$ at large $\rho$,
\begin{eqnarray}\label{Expandingoffrho}
f(\rho)=\frac{\rho^2}{\tilde L^2}-1+\mu \Big(\sum_{p=1}^{[\frac{d}{2}]}p\lambda_p f_\infty^{p-1} \Big)^{-1}\frac{1}{\rho^{d-2}}+...,
\end{eqnarray}
the constant $N$ in (\ref{BulkGeometry}) should be $\tilde L$. Plugging  (\ref{KeyFormula}) into (\ref{Lovelockderivative})  we get the solution of the bulk geometry with the period $\tau\sim \tau +2n\pi$, with the root of the following algebraic equation
\begin{eqnarray}\label{Alegbraequation}
\frac{2}{n\tilde L}\sum_{p=0}^{[\frac{d}{2}]}p\lambda_p \rho_{H}^{d-2p}L^{2p-2}+\sum_{p=0}^{[\frac{d}{2}]}(d-2p)\lambda_p \rho_{H}^{d-2p-1}L^{2p-2}=0.
\end{eqnarray}
It is obvious that the solution for $n=1$ is $\rho_{H}=\tilde L$ by using eq.(\ref{ConstraintOflambda}), which is the location of
the horizon of the pure AdS Lovelock solution in the hyperbolic foliation.\\

The second term in (\ref{Lovelock}) is the boundary term at the infinity $\rho=\rho_{\infty}$, and contains the curvature term $\mathcal R$ and  extrinsic curvature term $K$ at the boundary. This would make the variation work well and renormalize the action. These terms, also referred to as the Kounterterms, are discussed in \cite{Olea:2006vd}\cite{Kofinas:2006hr}\cite{Kofinas:2007ns}, see appendix B for more details. The action is different for even and odd dimensions, we should discuss them separately. We just quote the renormalized action for even dimension (see appendix B),
\begin{eqnarray}\label{ActionofLovelock}
I(n)=\frac{V_\Sigma}{16\pi G}\Big[&-&4\pi\sum_{p=1}^{[\frac{d}{2}]}\frac{d-1}{d-2p+1}p\lambda_p\rho_H^{d-2p+1}L^{2p-2}\nonumber \\
&-&2n\pi \tilde L (d-1)\mu +\frac{4\pi (d-1)!!^2}{d}\sum_{p=1}^{[\frac{d}{2}]}\frac{d-1}{d-2p+1}p \lambda_p \tilde L^{d-2p+1}L^{2p-2}\Big],
\end{eqnarray}
where $\mu$ is given by (\ref{mu}) and $\rho_{H}$ is the root of (\ref{Alegbraequation}). Taking $\rho_H=\tilde L$ and $n=1$ we get action $I(1)$.
The entanglement \ren entropy (\ref{RenyiFormula}) is
\begin{eqnarray}
S_n=\frac{n V_\Sigma}{4(n-1)G}\Big[&&\frac{1}{n}\sum_{p=1}^{[\frac{d}{2}]}\frac{d-1}{d-2p+1}p\lambda_p\rho_H^{d-2p+1}L^{2p-2}\nonumber \\
&&-\tilde L^{d-1}\sum_{p=1}^{[\frac{d}{2}]}p\lambda_p f_\infty^{2p-2}+\frac{1}{2}(d-1)\mu\Big].
\end{eqnarray}
For the odd dimensions the renormalized action does not contain the third term, which is related to the vacuum energy, in the square bracket of (\ref{ActionofLovelock}), see appendix B.
This term does not contribute to the entanglement \ren entropy, so the result in odd dimensions is the same as in even dimensions.\\

The Gauss-Bonnet gravity ($d\ge 4$) is a special case of  Lovelock gravity theory with $\lambda_0=1$, $\lambda_1=-1$, $\lambda_2\equiv \lambda$ and other coefficients being zero. The entanglement \ren entropy is
\begin{eqnarray}
S_n&=&\frac{n V_\Sigma \tilde L^{d-1}}{8(n-1)G}\Big[\frac{\left(\frac{d \rho _H^4}{f_{\infty }}+(d-4) \lambda  f_{\infty }-(d-2) \rho _H^2\right) \left(\frac{2 (d-1) \lambda  f_{\infty } \rho _H^{d-4}}{d-3}-\rho _H^{d-2}\right)}{\rho _H^2-2 \lambda  f_{\infty }}\nonumber\\
&&-\frac{4 (d-1) \lambda  f_{\infty }}{d-3}+2
+(d-1) \left(\lambda  f_{\infty } \rho _H^{d-4}+\frac{\rho _H^d}{f_{\infty }}-\rho _H^{d-2}\right)\Big],
\end{eqnarray}
the result is the same as in \cite{Hung:2011nu}. In the limit $\lambda\to 0$ we get the entanglement \ren entropy
for Einstein gravity in arbitrary dimensions
\begin{eqnarray}
S_n= \frac{n V_\Sigma L^{d-1}}{8(n-1)G}(2-\rho_H^d-\rho_H^{d-2}).
\end{eqnarray}

\subsection{Relation between generalized entropy method and previous approach}
In \cite{Casini:2011kv} the authors use a different approach and finally get the holographic entanglement entropy formula (\ref{HolographicFormula}).
The construction of \cite{Casini:2011kv} relies on conformal transformation which maps the causal development $\mathcal D $ of the ball to the new
geometry $R\times H^{d-1}$. Using the algebraic approach to quantum field theory they obtain the result that the reduced density matrix of the ball is related
to the thermal density matrix on the geometry $R\times H^{d-1}$. The entropy of thermal density state can be calculated by AdS/CFT correspondence, which relates this thermal state to the bulk topological black hole.

The generalized gravitational entropy is similar as the above statement. In section \uppercase\expandafter{\romannumeral2}  transformation (\ref{Transformation}) is the same as the one in
\cite{Casini:2011kv} in the Euclidean version. Actually the generalized gravitational entropy method does not need to prove that the density matrix on $S^1\times H^{d-1}$ is
thermal. A conformal transformation maps our spatial subregion $A$ and its complementary $\bar A$ into the surface $\tau=0$ and $\tau=\pi$ in the new spacetime respectively.  The DOFs on $\tau=0$ and
$\tau=\pi$ represent the whole system. Coordinate $\tau$ in the new spacetime is periodic and does not shrink to zero at the entangling surface. So the ``replica trick'' would produce
the smooth geometry\footnote{The idea of generalized gravitational entropy is analogous to the paper \cite{Holzhey:1994we}, in which the authors find the conformal transformation mapping the 2-dimensional Euclidean spacetime to strip, and the internal and its complementary part are mapped to two boundary of the strip.}. By using AdS/CFT correspondence we can calculate the density matrix. Actually this property also appears in \cite{Barvinsky:1994jca}\cite{Maldacena:2001kr}, in which the authors discuss the entanglement of the eternal black hole.  \\

The wave function of the system on $S^1\times H^{d-1}$ can be written as the path integral of fields in the interval $0\le\tau\le \pi$ with the boundary conditions $\phi(\tau=0)=\phi_1$ and $\phi(\tau=\pi)=\phi_2$. After tracing over the DOFs of the subregion $\bar A$, i.e., the surface $\tau=\pi$ in the new geometry, the reduced density matrix $\rho_A$ is equal to the partition function at temperature $T=\frac{1}{2\pi}$ derived in \cite{Casini:2011kv}. The n-th power of $\rho_A$ is still
a thermal state with  temperature  $T/n$. This just tells  the fact that the thermal density matrix can arise from entanglement, see \cite{Maldacena:2001kr}. The two approaches for spheric entangling surface can be viewed as the same phenomenon in different perspectives. But the latter is easily  generalized to arbitrary entangling surface.
Our calculation of the entanglement \ren entropy in this paper is similar   to \cite{Hung:2011nu}, as expected.
The approach in \cite{Casini:2011kv} can be seen as evidence for the generalized gravitational entropy method.

\section{Entanglement Entropy beyond Simple case}
Our discussion above is about a sphere and an infinite plane, in both cases the
dual bulk geometry can be found. For a general entangling surface it is expected that the geometry wll
depend on Euclidean time $\tau$, it seems impossible to construct the analytical
solution of the bulk geometry. In this section we use the argument of generalized entropy and
try to get some results for an entangling surface beyond the simple cases. \\

Let us first consider the entangling surface to be an infinite stripe in d-dimensional flat spacetime.
The subregion $A$ is defined as $x_0=0$, $0<x_1<l$, $-l_0<x_i<l_0$$(i=2,...,d-1)$, where $l_0$ is infinite.
There are two separated entangling surfaces located at $x_1=0$ and $x_1=l$. Unlike the infinite plane that
we discussed in section \uppercase\expandafter{\romannumeral2} it seems impossible to map the entangling area into a region and keep the
metric to be independent of the new time coordinate $\tau$. The dependence on time $\tau$ is expected for
a general configuration when we try to make a conformal transformation, see \cite{Lewkowycz:2013nqa}. We expect that the boundary defines the bulk geometry and
the $\tau$ circle shrinks to zero in the bulk. There are surfaces where
without knowing the exact form of the bulk geometry we can also get the leading contribution.\\

The metric of the d-dimensional spacetime can be written as
\begin{eqnarray}\label{MetricStripe}
ds^2=dzd\bar z+\sum_{i=2}^{d-1}dx_i^2,
\end{eqnarray}
where $z=ix_0+x_1$. We consider the following transformation,
\begin{eqnarray}\label{TransfromationStripe}
\frac{z-l}{z}=e^{-\omega},
\end{eqnarray}
and define the new coordinate $\omega=i\tau+u$. The metric (\ref{MetricStripe}) becomes
\begin{eqnarray}\label{StripeMetric}
ds^2=\Omega^2\Big(d\tau^2+du^2+4\frac{(\cosh u-\cos\tau)^2}{l^2}\sum_{i=2}^{d-1}dx_i^2\Big),
\end{eqnarray}
where $\Omega^2=l^{-2}z\bar z (z-l)(\bar z-l)$. The subregion $A$ is mapped to $\tau=\pi$ and $-\infty<u<+\infty$, $u=-\infty (+\infty)$
corresponds to the entangling surface $x_1=0(l)$. Its complementary part $\bar A$ is mapped to $\tau=0$. This is
what we expect. Actually we may find many different transformations to get the desired result. Another argument for
the above transformation (\ref{TransfromationStripe}) is that we would obtain the transformation for the infinite plane (\ref{TransformationPlane}) in the limit $l\to\infty$.
In fact in the limit $l\to \infty$, $\Omega^2 \to z\bar z\equiv x_0^2+x_1^2$, $\Omega^2 du^2\to (d\sqrt{z\bar z})^2$. The limit $l\to \infty$ means we only have one entangling plane located at $x_1=0$. \\

Although unable to construct the dual bulk geometry with a ``time'' dependent boundary, we can analyse its asymptotical behavior.
In the limit $|u|\to \infty$, metric (\ref{StripeMetric}) is asymptotically conformal to
\begin{eqnarray}\label{AsymptoticalMetric}
 ds^2=d\tau^2+du^2+\frac{e^{\pm 2u}}{l^2}\sum_{i=2}^{d-1}dx_i^2,
 \end{eqnarray}
where $\pm$ depends on limit $u\to -\infty(+\infty)$. Near the region $|u|\sim \infty$ we define the new coordinate $r=le^{\pm u}$,
metric (\ref{AsymptoticalMetric}) changes to
\begin{eqnarray}\label{PlaneSimilar}
ds^2=d\tau^2+\frac{dr^2+\sum_{i=2}^{d-1}dx_i^2}{r^2},
\end{eqnarray}
 This is just metric (\ref{TransformationPlane}) in section 2.1. We expect the dual geometry of (\ref{StripeMetric}) is asymptotic to (\ref{DualGeoofPlane}), which is the dual geometry of boundary metric (\ref{PlaneSimilar}). The leading contribution to the entanglement entropy is
\begin{eqnarray}\label{AreaLawStripe}
S_A\sim \int dr dx_2...dx_{d-1} \frac{1}{r^{d-1}}\sim \frac{A_{ent}}{\delta^{d-2}},
\end{eqnarray}
where $A_{ent}\equiv l_0^{d-2}$ denotes the area of the entangling surface, $\delta$ is the cut off to regularize the result. This is required by the area law of entanglement entropy.
Notice that we have two entangling surfaces here, their leading contributions both have the form of (\ref{AreaLawStripe}). The contribution
is from the region $|u|\sim \infty$, near the entangling surfaces. This result confirms the fact that the leading contribution of the entanglement entropy is due to the short distance coupling near the entangling surface. From the transformation we observe that $z\to 0$,  the entangling surface, in the limit $u\to \infty$. The new coordinate $r=le^{-u} \to \sqrt{z\bar z}=\sqrt{x_0^2+x_1^2}$, consistent with the coordinate $r$ that is defined in section 2.1. This implies that  transformation (\ref{TransfromationStripe})
reduces to simple transformation of the infinite plane (\ref{TransformationPlane}). This is natural because the entangling surfaces locally look like an infinite plane.\\
For general entangling surface the metric can be written in Gauss normal coordinates
\begin{eqnarray}\label{MetricArbitrary}
ds^2=dt^2+dy^2+g_{ij}(y,x)dx^idx^j,\quad i,j=1,...,d-2,
\end{eqnarray}
where $y$ is the coordinate normal to the entangling surface, $x_i$ are the ones tangent to the surface. The
metric $g_{ij}(y,x)=g_{ij}(0,x)+2yK_{ij}+O(y^2)$ near the surface. The entangling surface is defined by $t=0$
and $y=0$. The subregion $A$ $(y<0)$ and its complementary part $\bar A$ are
mapped to a $\tau=0$ and $\tau=\pi$ by a suitable transformation $U$, where $\tau$ is periodic and encircles  the entangling surface. The new geometry is
generally  time dependent. Our discussion in the
above example can be generalized to an arbitrary entangling surface.  Metric (\ref{MetricArbitrary}) near the entangling surface is
\begin{eqnarray}
 ds^2=dt^2+dy^2+\delta_{ij}d\xi^id\xi^j+..., \quad i,j=1,...,d-2,
\end{eqnarray}
where new coordinates $\xi_i$ satisfy $\delta_{ij}d\xi^id\xi^j=g(0,x)dx^idx^j$, ``...'' are terms are of higher orders. As we have mentioned the local property of the
entangling surface is same as an infinite plane. We expect the metric (\ref{MetricArbitrary}) after transformation $U$ would reduce to
\begin{eqnarray}\label{InfintePlaneTrans}
ds^2=\Omega^2(d\tau^2+\frac{dr^2+\delta_{ij}d\xi^id\xi^j}{r^2}),
\end{eqnarray}
near the entangling surface, where $y=r\cos\tau$, $t=r\sin\tau$, and $\Omega=r^2$, see the above example. And the dual bulk geometry would also have an
asymptotical behavior as the dual geometry of (\ref{InfintePlaneTrans}) locally. We can glue the different patches smoothly and choose
a cut-off $r\sim \delta$, the leading contribution of the entanglement entropy would be
\begin{eqnarray}\label{AreaLaw}
S_{leading}&\sim& \int_{r\sim 0}drd\xi_1d\xi_2...d\xi_{d-2}\frac{1}{r^{d-1}}
\nonumber \\
&=&\int_{r\sim 0}dr dx_1...dx_{d-2}\frac{1}{r^{d-1}}\sqrt{Det g_{ij}(0,x)}\sim\frac{A_{ent}}{\delta^{d-2}},
\end{eqnarray}
where $r\sim 0$ means the integral region is near the entangling surface and along the entangling surface, and we use
$|\frac{\partial \xi_i}{\partial x_j}|=\sqrt{Det g_{ij}(0,x)}$. (\ref{AreaLaw}) is the area law of the entanglement entropy\cite{Srednicki:1993im}\cite{Eisert:2008ur}.

\section{Discussion and Conclusion}
Our above discussion is mainly about the entanglement property of a spatial subregion $A$ on the boundary
of the bulk geometry. The generalized gravitational entropy is seen as the the entanglement entropy of
more general system, the holographic entanglement entropy is only an example by the $AdS/CFT$ correspondence.
For gravity theory with a
dual boundary theory the density matrix $\rho$ in the bulk has a well defined map in the boundary field theory. This motivates
a similar explanation for gravity theory without knowing the dual boundary field theory.
For a general quantum theory we should know the wave function of the system to define the entanglement
property of subregion in it. We follow the proposal of \cite{Barvinsky:1994jca}, where the authors consider the eternal version
of a Schwarzchild black hole with mass $M$ in 4 dimension. The wave function of the black hole is the Euclidean path
integral over field configurations $\phi$ on the region $-2\pi M<\tau<2\pi M$, where $\tau$ is the Euclidean time with the
period  $8\pi M$. $\phi(\tau=0)=\phi_1$ and $\phi(\tau=4\pi M)=\phi_2$ are the boundary conditions of the
wave function. The boundary is the Cauchy surface with the topology $R\times S^2$. As was shown in \cite{Barvinsky:1994jca},
 boundaries $\tau=0$ and $\tau=4\pi M$ represent the inside and the outside DOFs respectively.
 The wave function can be written as a path integral
\begin{eqnarray}\label{WaveFuncofBlackhole}
\Phi[\phi_1,\phi_2]=\int_{\phi(\tau=0)=\phi_1,\phi(\tau=4\pi M)=\phi_2}D\phi e^{-I_E},
\end{eqnarray}
where $I_E$ is the Euclidean action including gravity and matter. With the wave function we can define the
reduced  density matrix $\rho$ by tracing over DOFs inside the horizon, a path integral on the
 $\tau$ circle with a cut at $\tau=8\pi M$. The n-th power of $\rho$ is given by the path integral on another
 spacetime with  $\tau \sim \tau +8n\pi M$,  similar to the generalized gravitational entropy.
 But in the new spacetime there is a conical singularity on the surface where the $\tau$ circle shrinks to zero \cite{Solodukhin S N}.
 This seems to contradict to the argument of the generalized gravitational entropy. But in fact we may consider the orbifold of the smooth bulk solution $M_n$ with $\tau\sim\tau+2n\pi$ $M_n/Z_n$, then there is a conical defect $8\pi M(1-1/n)\equiv 8\pi M\epsilon$ at the horizon. We can calculate the on-shell action $\hat I$ without considering the effect of the conical defect, then $I(n)=n\hat I$. Actually $\partial_\epsilon \hat I$ is equal to the variance of the action with an opening angle $8\pi M(1+\epsilon)$ by $\epsilon$, see the paper\cite{Lewkowycz:2013nqa} \cite{Dong:2013qoa}. We can use both methods to calculate entropy. But the generalized gravitational entropy offers a physical explanation. The calculation is simple in the semiclassical approximation . The smooth spacetime with period
 $\tau\sim\tau+8n\pi M$ is the Schwarzchild solution with  parameter $M$  replaced by  $n M$. The renormalized on-shell action is
 $I(n)=-\frac{4\pi nM^2}{G} $\footnote{The renormalization procedure is the same as in \cite{Gibbons}}, so the entanglement \ren entropy is  $S_n=\frac{4nM^2}{G}$.
 We find the entanglement entropy $S=\lim_{n\to 1}S_n=\frac{A}{4G}$. Actually
 the on-shell action of the smooth spacetime and the one with a conical singularity are equal in the limit $n\to 1$. The entanglement
 \ren entropy due to the generalized gravitational entropy also has a simple and nice relation with the thermodynamical ensemble of the
 black hole, we can slightly modify the \ren entropy formula (\ref{RenyiCalculation}) as in \cite{Baez}
 \begin{eqnarray}
 S_n=\frac{TI(n)-T_0I(1)}{T-T_0}=-\frac{F(T)-F(T_0)}{T-T_0},
 \end{eqnarray}
where $T_0$ is the temperature of the black hole with parameter $M$, $F(T)=-TI(n)$ is the free energy of the system at temperature $T$, we define the new temperature $T=T_0/n$. We observe that $T$ is
just the black hole temperature with the period $\tau\sim \tau+8n\pi M$. In the limit $n\to 1$, $S_n\to S=-\frac{\partial F}{\partial T}$, which is the thermal relation between the free energy and the entropy of the system. Our discussion here is valid only in the semiclassical approximation, it is worth to study whether  the on-shell
generalized gravitational entropy method gives the same answer as in  the
off-shell approach at the quantum level when quantum matter appears.\\

Another interesting thing concerning  the entanglement entropy is the the variance of entanglement entropy for excited state that is discussed in many paper recently, e.g., \cite{Bhattacharya:2012mi}-\cite{He:2013rsa}. It's interesting to consider how to understand the variance of the entanglement entropy by the generalized gravitational entropy.\\
In this paper we use the generalized gravitational entropy method to study the \ren entropy and entanglement entropy for a subregion in
conformal field theory. For the simple sphere case we find the transformation mapping the region to desired geometry with a bulk dual. By AdS/CFT correspondence we calculate the entanglement \ren entropy of the sphere for  Einstein gravity. We find that the approach has a close
relation with the previous one \cite{Casini:2011kv} and confirm the relation between the reduced  and the thermal states for the field theory. We also study the
entangling surface beyond the simplest case. Although unable to construct the dual bulk geometry for a general case, one can get some information
about the entanglement entropy from the local property near the entangling surface. And we derive the famous area law of entanglement entropy by this method.

\section{Acknowledgement}
We wish to thank Song He and Rong-xin Miao for useful discussions on this topic.

\appendix
\section*{APPENDIX}
\section{Foliation of $AdS_{d+1}$}
In this section we want to complete the last step of the proof for the infinite plane and the sphere, i.e.,
to show that the horizon, which is a minimal surface, is homologous to the entangling surface.
The Euclidean version of $AdS_{d+1}$ geometry with radius $L=1$ can be described by the following surface \cite{Casini:2011kv}
\begin{eqnarray}\label{Surface1}
-y_{-1}^2+\sum_{i=0}^{d-1}y_i^2+y_{d}^2=-1
\end{eqnarray}
embedded in $R^{1,d+1}$ with metric
\begin{eqnarray}\label{EmbeddeMetric}
ds^2=-y_{-1}^2+\sum_{i=0}^{d-1}dy_i^2+dy_{d}^2
\end{eqnarray}
We get the standard Poincar\'e patch of $AdS_{d+1}$ space with the following foliation
\begin{eqnarray}\label{FolitaionPoincare}
y_{-1}+y_{d}=\frac{1}{z},\quad \quad y_i=\frac{x_i}{z},\quad with\  i=0,1,..., d-1.
\end{eqnarray}
The metric in these coordinates is
\begin{eqnarray}
ds^2=\frac{1}{z^2}\Big(dz^2+\sum_{i=0}^{d-1}dx_i^2\Big).
\end{eqnarray}
The asymptotic boundary by taking the limit $z\to 0$  is conformal to $R^d$.
We assume our entangling surface is $x=0$. \\
The Euclidean version of $AdS_{d+1}$ can also be foliated by surfaces with the geometry $R\times H^{d-1}$ as follows.
\begin{eqnarray}\label{DualGeoofPlane}
ds^2=\frac{d\rho^2}{\rho^2-1}+(\rho^2-1)d\tau^2+\rho^2\Big(\frac{dr^2+\sum_{i=1}^{d-2}dx_i^2}{r^2}\Big).
\end{eqnarray}
We can get above metric by embedding surface $(\ref{Surface1})$ into $R^{1,d+1}$ with the following
coordinate transformation
\begin{eqnarray}
y_{-1}+y_{d}=\frac{\rho}{r}, \quad \quad y_{0}=\bar\rho \sin\tau,\quad \quad y_{1}=\bar\rho \cos\tau, \quad \quad
y_i=\frac{\rho x_i}{r},\quad with\  i=2,...,d-1, \nonumber
\end{eqnarray}
and the constraint
\begin{eqnarray}
-y_{-1}^2+\sum_{i=2}^{d}y_{i}^2=-\rho^2,
\end{eqnarray}
where $\bar \rho=\sqrt{\rho^2-1}$. The horizon of the topological black hole sits at $\rho=1$ ($\tau$ is arbitrary).
$y_{0}=0$ and $y_{1}=0$ indicate that the horizon is really homologous to the entangling surface $x_{0}=0$, $x_{1}=0$.\\

Our construction of the bulk geometry is actually the same as the approach in \cite{Casini:2011kv}. Unlike the infinite plane we foliate
the Euclidean $AdS_{d+1}$ in a new way to be consistent with boundary (\ref{BoundarySphere}). We list the foliation
as follows.
\begin{eqnarray}
y_{-1}=\rho \cosh u,\quad  y_{1}=\rho \sinh u \cos \phi_{1}, \quad y_{2}=\rho \sinh u \sin\phi_{1} \cos\phi_{2},...,\nonumber \\
y_{d-1}=\rho \sinh u \sin\phi_{1} \sin\phi_{2}...\cos\phi_{d-2},\quad y_{0}=\bar \rho \sin \tau,\quad y_d=\bar\rho\cos\tau,
\end{eqnarray}
where $\bar\rho=\sqrt{\rho^2-1}$.
Then the metric is
\begin{eqnarray}
ds^2=\frac{d\rho^2}{\rho^2-1}+(\rho^2-1)d\tau^2+\rho^2\Big(du^2+\sinh^2ud\Omega^2_{d-2}\Big).
\end{eqnarray}
The horizon is $\rho=1$($\tau$ is arbitrary). According to (\ref{FolitaionPoincare}) we get
\begin{eqnarray}
\frac{1}{z}=\cosh u\quad and \quad r^2=\sum_{i=1}^dx_i=z^2\sinh^2u.
\end{eqnarray}
The surface would reach  the boundary of $AdS_{d+1}$ as $z\to 0$, or equally $u\to \infty$. We get
$r\to 1$ in this limit, which is just what we want to show. The horizon satisfies the constraint equation
$r^2+z^2=1$, which is the minimal surface derived from (\ref{HolographicFormula}) \cite{Ryu:2006ef}\cite{Nishioka:2009un}.

\section{The Euclidean action}
\subsection{Einstein gravity }
Let us discuss the Einstein gravity first. Using the solution of equation of motion (\ref{EinsteinSoltion}), we get
$R=-d(d+1)$
\begin{eqnarray}
I_{bulk}=\frac{n V_{\Sigma}}{4G}(\rho_\infty^d-\rho_H^d),
\end{eqnarray}
\begin{eqnarray}
I_{surface}=-\frac{n V_\Sigma}{4G}\Big(d \rho_\infty^d-(d-1)\rho_\infty^{d-2}+\frac{d}{2}(\rho_H^{d-2}-\rho_H^d)\Big)+O(\frac{1}{\rho_\infty^2})
\end{eqnarray}
We observe that both two terms contain divergence. The possible counter terms  to be added to the action is
discussed in many literatures, for example, \cite{Emparan:1999pm}\cite{de Haro:2000xn}. We quote the form  of counter term for $d=2$, $d=4$, $d=5$,
\begin{eqnarray}
d=2\quad \quad I_{ct}=\frac{1}{8\pi G}\int_{\partial M}\sqrt{h}=\frac{n V_\Sigma}{4G}(\rho_\infty^2-\frac{\rho_H^2}{2})+O(\frac{1}{\rho_\infty^2}),
\end{eqnarray}
\begin{eqnarray}
d=3\quad \quad I_{ct}=\frac{1}{8\pi G}\int_{\partial M}\sqrt{h}(2+\frac{\mathcal R}{2})=2\rho_\infty^3-2\rho_\infty+(\rho_H-\rho_H^3)+O(\frac{1}{\rho_\infty}),
\end{eqnarray}
\begin{eqnarray}
d=4 \quad \quad I_{ct}=\frac{1}{8\pi G}\int_{\partial M}\sqrt{h}(3+\frac{\mathcal R}{4})=3\rho_\infty^4-3\rho_\infty^2+\frac{3}{8}(1+4\rho_H^2-4\rho_H^4)+O(\frac{1}{\rho_\infty^2}),
\end{eqnarray}
\begin{eqnarray}
d=5 \quad \quad I_{ct}&=&\frac{1}{8\pi G}\int_{\partial M}\sqrt{h}(4+\frac{\mathcal R}{6}+\frac{1}{18}(\mathcal R_{ab}\mathcal R^{ab}-\frac{5}{16}\mathcal R^2))\nonumber \\
&=& 4\rho_\infty^5-4\rho_\infty^3+2\rho_H^3-2\rho_H^5+O(\frac{1}{\rho_\infty}),
\end{eqnarray}
where $\mathcal R$ and $\mathcal R_{ab}$ are  Ricci scalar and Ricci tensor of the induced metric.
The renormalized Euclidean action of $I(n)$ is
\begin{eqnarray}
d=2 \quad I(n)=-\frac{n V_{\Sigma}}{8G}(\rho_H^2+2)+O(\frac{1}{\rho_\infty^2}),
\end{eqnarray}
\begin{eqnarray}
d=3 \quad I(n)=-\frac{n V_{\Sigma}}{8G}(\rho_H^3+\rho_H)+O(\frac{1}{\rho_\infty}),
\end{eqnarray}
\begin{eqnarray}
d=4 \quad I(n)=-\frac{n V_{\Sigma}}{8G}(\rho_H^2+\rho_H^4-\frac{3}{4})+O(\frac{1}{\rho_\infty^2}),
\end{eqnarray}
\begin{eqnarray}
d=5 \quad I(n)=-\frac{n V_{\Sigma}}{8G}(\rho_H^3+\rho_H^5)+O(\frac{1}{\rho_\infty}),
\end{eqnarray}
The counter term also contributes to the entanglement \ren entropy as we can see in the above calculation,
this kind contribution also appears in the quantum correction \cite{Faulkner:2013ana}.\\

\subsection{Lovelock gravity for arbitrary dimension}
In this subsection we deal with the renormalized action of Lovelock gravity, we will separately discuss odd and even dimensional cases.
We would work in Euclidean spacetime. Direct calculation shows that the first term in (\ref{Lovelock})
is
\begin{eqnarray}\label{LovelockBulk}
I_{bulk}=\frac{\beta V_\Sigma}{16 \pi G}\sum_{p=1}^{[\frac{d}{2}]}\frac{(d-1)L^{2p-2}}{d-2p+1}p\lambda_p\Big[\rho^{d+1-2p}f'{\rho}\Big(1+f{\rho})^{p-1}\Big)|^{\rho_\infty}_{\rho_H},
\end{eqnarray}
where we denote $\beta=2n\pi \tilde L$. The boundary term in (\ref{Lovelock}) is constructed in \cite{Kofinas:2007ns} for Lovelock AdS gravity.
the general form of $B_{d}$, referred to as Kounterterms, in odd dimensions is
\begin{eqnarray}
B_d=d\sqrt{-h}\int_0^1dt\int^1_0ds&& \delta^{[i_1...i_{d-1}]}_{[j_1...j_{d-1}]}K^{j_1}_{i_{1}}\Big(\frac{1}{2}\mathcal R^{j_2 j_3}_{i_2 i_3}-t^2K^{j_2}_{i_{2}}K^{j_3}_{i_{3}}+\frac{s^2}{\tilde L^2}\delta_{i_2}^{j_2}\delta^{j_3}_{i3}\Big)...\times\nonumber \\
&&\Big(\frac{1}{2}\mathcal R^{j_{d-2} j_{d-1}}{}_{i_{d-2} i_{d-1}}-t^2K^{j_{d-2}}_{i_{d-2}}K^{j_{d-1}}_{i_{d-1}}+\frac{s^2}{\tilde L^2}\delta_{i_{d-2}}^{j_{d-2}}\delta^{j_{d-1}}_{i_{d-1}}\Big)d^{d}x,
\end{eqnarray}
where $\mathcal R_{ij}^{kl}$ and $K_{ij}$ are the intrinsic and the extrinsic curvatures of the induced hyperspace $\rho=\rho_{\infty}$. In \cite{Kofinas:2007ns} the authors show the surface term makes the variance method work, if the coefficient
\begin{eqnarray}\label{CD}
c_d=\frac{1}{8d G}\Big[\int_0^1dt(t^2-1)^{n-1} \Big]^{-1} \sum_{p}^{[\frac{d}{2}]}\frac{p\tilde L^{2d-2}}{(d-2p+1)(d-2)!}\lambda_p f_{\infty}^{2p-2}.
\end{eqnarray}
The result of the surface term is
\begin{eqnarray}\label{Surface}
I_{surface}&=&-\frac{c_d d!V_{\Sigma}}{2}\beta \Big[f'(\rho)\int_0^1dt (-1-f(\rho)+\frac{t^2 \rho^2}{\tilde L^2})^{n-1}\nonumber \\
&+&(2f(\rho)-\rho f'(\rho) )\int_0^1dt t(-1-t^2f(\rho)+\frac{t^2 \rho^2}{\tilde L^2})^{n-1}\Big]|_{\rho=\rho_\infty}.
\end{eqnarray}
Taking (\ref{Expandingoffrho}) and (\ref{CD}) into (\ref{Surface}), we get the surface term in the form of a series in $\rho$, and one could find the infinity term of $I_{bulk}$ and $I_{surface}$ are canceled. The renormalized action of Lovelock gravity for odd dimension is
\begin{eqnarray}\label{ODD}
I_{odd}=\frac{V_\Sigma}{16\pi G}\Big[&-&4\pi\sum_{p=1}^{[\frac{d}{2}]}\frac{d-1}{d-2p+1}p\lambda_p\rho_H^{d-2p+1}L^{2p-2}\nonumber \\
&-&2n\pi \tilde L (d-1)\mu +\frac{4\pi (d-1)!!^2}{d}\sum_{p=1}^{[\frac{d}{2}]}\frac{d-1}{d-2p+1}p \lambda_p \tilde L^{d-2p+1}L^{2p-2}\Big].
\end{eqnarray}
The method to deal with an even dimensional case is similar. The result of the renormalized action is
\begin{eqnarray}
 I_{even}=\frac{V_\Sigma}{16\pi G}\Big[-4\pi\sum_{p=1}^{[\frac{d}{2}]}\frac{d-1}{d-2p+1}p\lambda_p\rho_H^{d-2p+1}L^{2p-2}
-2n\pi \tilde L (d-1)\mu \Big],
\end{eqnarray}
it is obvious that the third term in (\ref{ODD}) does not appear in even dimensions.


\begin{thebibliography}{99}
%\cite{Ryu:2006bv}
\bibitem{Ryu:2006bv}
  S.~Ryu and T.~Takayanagi,
  ``Holographic derivation of entanglement entropy from AdS/CFT,''
  Phys.\ Rev.\ Lett.\  {\bf 96}, 181602 (2006)
  [hep-th/0603001].

%\cite{Fursaev:2006ih}
\bibitem{Fursaev:2006ih}
  D.~V.~Fursaev,
  ``Proof of the holographic formula for entanglement entropy,''
  JHEP {\bf 0609}, 018 (2006)
  [hep-th/0606184].

  %\cite{Headrick:2010zt}
\bibitem{Headrick:2010zt}
  M.~Headrick,
  ``Entanglement Renyi entropies in holographic theories,''
  Phys.\ Rev.\ D {\bf 82}, 126010 (2010)
  [arXiv:1006.0047 [hep-th]].

  \bibitem{Casini:2011kv}
  H.~Casini, M.~Huerta and R.~C.~Myers,
  ``Towards a derivation of holographic entanglement entropy,''
  JHEP {\bf 1105}, 036 (2011)
  [arXiv:1102.0440 [hep-th]].

  %\cite{Hung:2011nu}
\bibitem{Hung:2011nu}
  L.~-Y.~Hung, R.~C.~Myers, M.~Smolkin and A.~Yale,
  ``Holographic Calculations of Renyi Entropy,''
  JHEP {\bf 1112}, 047 (2011)
  [arXiv:1110.1084 [hep-th]].

%\cite{Lewkowycz:2013nqa}
\bibitem{Lewkowycz:2013nqa}
  A.~Lewkowycz and J.~Maldacena,
  ``Generalized gravitational entropy,''
  JHEP {\bf 1308}, 090 (2013)
  [arXiv:1304.4926 [hep-th]].

%\cite{Holzhey:1994we}
\bibitem{Holzhey:1994we}
  C.~Holzhey, F.~Larsen and F.~Wilczek,
  ``Geometric and renormalized entropy in conformal field theory,''
  Nucl.\ Phys.\ B {\bf 424}, 443 (1994)
  [hep-th/9403108].
%\cite{Barvinsky:1994jca}
\bibitem{Barvinsky:1994jca}
  A.~O.~Barvinsky, V.~P.~Frolov and A.~I.~Zelnikov,
  ``Wavefunction of a Black Hole and the Dynamical Origin of Entropy,''
  Phys.\ Rev.\ D {\bf 51}, 1741 (1995)
  [gr-qc/9404036].

  \bibitem{Solodukhin S N}
  Solodukhin S N. `` Entanglement Entropy of Black Holes,'' Living Rev. Rel, 2011, 14(8) [hep-th/1104.3712].

  \bibitem{Gibbons}
  Gibbons, G. W. and Stephen W. Hawking. ``Action integrals and partition functions in quantum gravity,'' Phys. Rev. D 15, 2752(1977).

  %\cite{Faulkner:2013ana}
\bibitem{Faulkner:2013ana}
  T.~Faulkner, A.~Lewkowycz and J.~Maldacena,
  ``Quantum corrections to holographic entanglement entropy,''
  arXiv:1307.2892 [hep-th].

  \bibitem{Baez}
  Baez J C. ``\ren entropy and free energy,'' arXiv:1102.2098[quant-ph].

  %\cite{Kofinas:2007ns}
\bibitem{Kofinas:2007ns}
  G.~Kofinas and R.~Olea,
  ``Universal regularization prescription for Lovelock AdS gravity,''
  JHEP {\bf 0711}, 069 (2007)
  [arXiv:0708.0782 [hep-th]]

  %\cite{Balasubramanian:1999re}
\bibitem{Balasubramanian:1999re}
  V.~Balasubramanian and P.~Kraus,
  ``A Stress tensor for Anti-de Sitter gravity,''
  Commun.\ Math.\ Phys.\  {\bf 208}, 413 (1999)
  [hep-th/9902121].

%\cite{Emparan:1999pm}
\bibitem{Emparan:1999pm}
  R.~Emparan, C.~V.~Johnson and R.~C.~Myers,
  ``Surface terms as counterterms in the AdS / CFT correspondence,''
  Phys.\ Rev.\ D {\bf 60}, 104001 (1999)
  [hep-th/9903238].

  %\cite{de Haro:2000xn}
\bibitem{de Haro:2000xn}
  S.~de Haro, S.~N.~Solodukhin and K.~Skenderis,
  ``Holographic reconstruction of space-time and renormalization in the AdS / CFT correspondence,''
  Commun.\ Math.\ Phys.\  {\bf 217}, 595 (2001)
  [hep-th/0002230].

  %\cite{Olea:2006vd}
\bibitem{Olea:2006vd}
  R.~Olea,
  ``Regularization of odd-dimensional AdS gravity: Kounterterms,''
  JHEP {\bf 0704}, 073 (2007)
  [hep-th/0610230].
  %\cite{Kofinas:2006hr}
\bibitem{Kofinas:2006hr}
  G.~Kofinas and R.~Olea,
  ``Vacuum energy in Einstein-Gauss-Bonnet AdS gravity,''
  Phys.\ Rev.\ D {\bf 74}, 084035 (2006)
  [hep-th/0606253].

  %\cite{Lovelock:1971yv}
\bibitem{Lovelock:1971yv}
  D.~Lovelock,
  ``The Einstein tensor and its generalizations,''
  J.\ Math.\ Phys.\  {\bf 12}, 498 (1971).

%\cite{Maldacena:2001kr}
\bibitem{Maldacena:2001kr}
  J.~M.~Maldacena,
  ``Eternal black holes in anti-de Sitter,''
  JHEP {\bf 0304} (2003) 021
  [hep-th/0106112].
%\cite{Eisert:2008ur}
\bibitem{Eisert:2008ur}
  J.~Eisert, M.~Cramer and M.~B.~Plenio,
  ``Area laws for the entanglement entropy - a review,''
  Rev.\ Mod.\ Phys.\  {\bf 82}, 277 (2010)
  [arXiv:0808.3773 [quant-ph]].
%\cite{Srednicki:1993im}
\bibitem{Srednicki:1993im}
  M.~Srednicki,
  ``Entropy and area,''
  Phys.\ Rev.\ Lett.\  {\bf 71}, 666 (1993)
  [hep-th/9303048].
  %\cite{Dong:2013qoa}
\bibitem{Dong:2013qoa}
  X.~Dong,
  ``Holographic Entanglement Entropy for General Higher Derivative Gravity,''
  arXiv:1310.5713 [hep-th].
\bibitem{Nishioka:2009un}
  T.~Nishioka, S.~Ryu and T.~Takayanagi,
  ``Holographic Entanglement Entropy: An Overview,''
  J.\ Phys.\ A {\bf 42}, 504008 (2009)
  [arXiv:0905.0932 [hep-th]].
  %\cite{Ryu:2006ef}
\bibitem{Ryu:2006ef}
  S.~Ryu and T.~Takayanagi,
  ``Aspects of Holographic Entanglement Entropy,''
  JHEP {\bf 0608}, 045 (2006)
  [hep-th/0605073].
  %\cite{Chen:2013qma}
\bibitem{Chen:2013qma}
  B.~Chen and J.~-j.~Zhang,
  ``Note on generalized gravitational entropy in Lovelock gravity,''
  JHEP {\bf 07}, 185 (2013)
  [arXiv:1305.6767 [hep-th]].
  %\cite{Bhattacharyya:2013jma}
\bibitem{Bhattacharyya:2013jma}
  A.~Bhattacharyya, A.~Kaviraj and A.~Sinha,
  ``Entanglement entropy in higher derivative holography,''
  JHEP {\bf 1308}, 012 (2013)
  [arXiv:1305.6694 [hep-th]].
  %\cite{Dong:2013qoa}
\bibitem{Dong:2013qoa}
  X.~Dong,
  ``Holographic Entanglement Entropy for General Higher Derivative Gravity,''
  arXiv:1310.5713 [hep-th].
  %\cite{Bhattacharya:2012mi}
\bibitem{Bhattacharya:2012mi}
  J.~Bhattacharya, M.~Nozaki, T.~Takayanagi and T.~Ugajin,
  ``Thermodynamical Property of Entanglement Entropy for Excited States,''
  Phys.\ Rev.\ Lett.\  {\bf 110}, no. 9, 091602 (2013)
  [arXiv:1212.1164].
  %\cite{Blanco:2013joa}
\bibitem{Blanco:2013joa}
  D.~D.~Blanco, H.~Casini, L.~-Y.~Hung and R.~C.~Myers,
  ``Relative Entropy and Holography,''
  JHEP {\bf 1308}, 060 (2013)
  [arXiv:1305.3182 [hep-th]].
  %\cite{Guo:2013aca}
\bibitem{Guo:2013aca}
  W.~-z.~Guo, S.~He and J.~Tao,
  ``Note on Entanglement Temperature for Low Thermal Excited States in Higher Derivative Gravity,''
  JHEP {\bf 1308}, 050 (2013)
  [arXiv:1305.2682 [hep-th]].
  %\cite{He:2013rsa}
\bibitem{He:2013rsa}
  S.~He, D.~Li and J.~-B.~Wu,
  ``Entanglement Temperature in Non-conformal Cases,''
  arXiv:1308.0819 [hep-th].
\end{thebibliography}
\end{document}